\documentclass[showpacs]{revtex4}%
\usepackage{amsfonts}
\usepackage{amsmath}
\usepackage{amssymb}
\usepackage{graphicx}%
\usepackage[colorlinks=true, citecolor=blue, urlcolor = blue, linkcolor= red, bookmarks=true]{hyperref}
\newcommand*{\car}{{\cal R}}
\begin{document}
\title{Radiating black holes in the novel 4D Einstein-Gauss-Bonnet gravity}
\author{Sushant G. Ghosh $^{a,\;b,}$} \email{sghosh2@jmi.ac.in, sgghosh@gmail.com}
\author{Sunil D. Maharaj $^{b,}$}\email{maharaj@ukzn.ac.za}
\affiliation{$^{a}$ Centre for Theoretical Physics, Jamia Millia Islamia, New Delhi 110 025, India}
\affiliation{$^{b}$ Astrophysics and Cosmology
Research Unit, School of Mathematics, Statistics and Computer Science, University of
KwaZulu-Natal, Private Bag 54001, Durban 4000, South Africa}

\date{\today}
\begin{abstract}
Recently  Glavan and Lin [Phys. Rev. Lett. 124, 081301 (2020)] formulated a novel Einstein-Gauss-Bonnet gravity in which the Gauss-Bonnet coupling has been rescaled as $\alpha/(D-4)$ and the $4D$ theory is defined as the limit $D \rightarrow 4$, which preserves the number degrees of freedom thereby free from the Ostrogradsky instability. We present  exact spherically symmetric nonstatic null dust solutions in the novel 4D Einstein-Gauss-Bonnet gravity that bypasses the Lovelock theorem.  Our solution represents radiating black holes  and regains, in the limit $\alpha \rightarrow 0$,   the famous  Vaidya  black hole  of general relativity (GR).  We discuss the horizon structure of  black hole solutions to find that the  three horizon-like loci that characterizes its structure, viz. $AH$, $EH$ and $ TLS $  have the relationship $r_{EH} < r_{AH} = r_{TLS}$.  The charged radiating black holes in the theory, generalizing Bonnor-Vaidya black holes, are also considered.  In particular our results, in the limit $\alpha \rightarrow 0$, reduced exactly  to  \emph{vis-$\grave{a}$-vis} $4D$  black holes of GR.
\end{abstract}


\maketitle

\section{INTRODUCTION}
Lovelock theory of gravity,  a natural  generalization of
Einstein's general relativity (GR), was introduced by David Lovelock
\cite{dll}, and is defined by the action
\begin{align}
\label{action}
{\cal I}=& \frac{1}{16 \pi G_D}\int  d^D x\sqrt{-g}\sum_{p=0}^{[D/2]}\alpha_{(p)}{\mathcal L}_{(p)}+I_{ matter},\\
 {\mathcal L}_{(p)}:=&\frac{1}{2^p}\delta^{\mu_1\cdots \mu_p\nu_1\cdots \nu_p}_{\rho_1\cdots \rho_p\sigma_1\cdots \sigma_p}R_{\mu_1\nu_1}^{\phantom{\mu_1}\phantom{\nu_1}\rho_1\sigma_1}\cdots R_{\mu_p\nu_p}^{\phantom{\mu_p}\phantom{\nu_p}\rho_p\sigma_p},
\end{align}
where  $\alpha_{(p)}$ is an arbitrary
constant with dimension $({\rm length})^{2(p-1)}$, and
$\mathcal{L}_{(p)}$ is the Euler density of a 2$p$-dimensional
manifold. The $\delta$ symbol denotes a totally antisymmetric
product of Kronecker deltas, normalized to take values $0$ and $\pm
1$~\cite{dll}, which are defined by
\begin{align}
\delta^{\mu_1\cdots \mu_p}_{\rho_1\cdots \rho_p}:=&p!\; \delta^{\mu_1}_{[\rho_1}\cdots \delta^{\mu_p}_{\rho_p]}.
\end{align}
The quantity $\alpha_{(0)}$ is related to the cosmological constant
$\Lambda$ by $\alpha_{(0)}=-2\Lambda$.  Among these terms, the
first term is the cosmological constant term ($\Lambda$), the second term is the
Einstein GR term ($\mathcal{R}$) term, and the third  term is the second-order Gauss-Bonnet contribution \cite{Lanczos:1938sf} given by
\begin{equation}
{\cal L}_{GB}=\car^2-4\car_{cd}\car^{cd}+\car_{cdef}\car^{cdef}.
\label{GB-Lagrangian}
\end{equation}
The special case in Lovelock gravity that received most significant attention is the so-called Einstein-Gauss-Bonnet gravity \cite{Lanczos:1938sf}, which naturally appears in the low energy effective action of heterotic string theory \cite{Gross}, and then the action (\ref{action}) reduces to
\begin{eqnarray}
{\cal I}=\frac{1}{16 \pi G_D}\int d^Dx \sqrt{-g}
\left[\car - 2\Lambda + \alpha {\cal L}_{GB}
\right].
\end{eqnarray}
The parameter $\alpha$ has dimension of (length)$^2$ and we  assume that $\alpha$ is positive. From the action, we obtain the following field equations \cite{Boulware:1985wk}
\begin{eqnarray}
{\cal G}_{ab}=\kappa^2{\cal T}_{ab}
-\Lambda g_{ab}+\frac{\alpha}{2}{\cal H}_{ab},
\label{field_eq}
\end{eqnarray}
where ${\cal G}_{ab}=\mathcal{R}_{ab}-\frac{1}{2}\mathcal{R}g_{ab}$ is the Einstein tensor,
${\cal T}_{ab}$ is the energy momentum tensor
of matter fields, and
\begin{eqnarray}
{\cal H}_{ab}:=
{\cal L}_{GB} g_{ab}
-4(\car\car_{ab}-2\car_{ac}\car^c_{~b}
-2\car_{acbd}\car^{cd}+\car_{acde}\car_b^{~cde}),
\end{eqnarray}
is the Lanczos tensor.  The  Einstein-Gauss-Bonnet gravity is  important in exploring many conceptual issues of gravity  in a more general setup, e.g., existence and uniqueness theorems, black hole thermodynamics, 
and entropy, their horizon properties.  

Let us consider maximally symmetric spacetimes in Einstein-Gauss-Bonnet gravity such that 
\begin{eqnarray}
\car_{abcd}={\cal K}\left(g_{ac}g_{bd}-g_{ad}g_{bc}\right),
\end{eqnarray}
with the curvature scale
\begin{eqnarray}
&&{\cal K}:=-\frac{1}{2\tilde\alpha}\left[1\pm
\sqrt{1+\frac{8\tilde\alpha\Lambda}{(D-1)(D-2)}} \right],\\
&&\quad\tilde \alpha:=(D-3)(D-4)\alpha,
\end{eqnarray}
that solves the field equations~(\ref{field_eq}).
Only in the ``$-$'' branch
we have ${\cal K}=\Lambda/3 $ 
in the limit of $\alpha\to 0$.   Although the ``$+$'' branch seems physically less important,
it is interesting to note that even when $\Lambda =0$  we have ${\cal K}<0$, namely, AdS spacetime.

The spherically symmetric static black hole  solution in  Einstein-Gauss-Bonnet theory was  first  obtained by Boulware and Deser \cite{Boulware:1985wk}  
\begin{eqnarray}
ds^2=-f(r)dt^2+\frac{1}{f(r)}dr^2+r^2\gamma_{ij}dx^idx^j,
\label{BH_metric}
\end{eqnarray}
where $\gamma_{ij}$ is a metric of the constant curvature $(D-2)$-space
parameterized by $k=0$ (flat space), $+1$ (sphere), $-1$ (hyperboloid),
and
\begin{eqnarray}
f(r)=k+
\frac{r^2}{2\tilde \alpha}
\left[1 \pm \sqrt{1+\frac{8\tilde \alpha\Lambda}{(D-1)(D-2)}
	+\frac{4\tilde \alpha M}{r^{D-1}}}
\right],
\end{eqnarray}
with $M$ being the constant of integration identified as gravitational mass.   
Later several interesting black hole solutions were obtained for various sources \cite{Wiltshire:1985us,egb, egb2} including the colored black holes solution of Yang-Mills-Dilaton-Gravity system in the presence of a Gauss-Bonnet term \cite{Kanti:1996gs}.  It was proved that  in the Einstein-scalar-Gauss-Bonnet theory with a coupling function the  no-hair theorems are easily evaded, and also lead to a large number of regular black hole solutions \cite{Antoniou:2017acq,Bakopoulos:2018nui,Antoniou:2017hxj} and  other regular black holes are solutions of Einstein-Gauss-Bonnet theory coupled to nonlinear electrodynamics \cite{Ghosh1:2018bxg}.     

Henceforth we will restrict our attention to the spherical case $k=1$. 
It turns out that the Gauss-Bonnet term is a topological invariant in $D=4$ as its contribution to all the components of Einstein's equation are in fact proportional to $(D-4)$ and one requires $D\geq 5$ for non-trivial gravitational dynamics. 
However by rescaling the Gauss-Bonnet coupling constant $\alpha$ as
\begin{equation}\label{scale}
\alpha \rightarrow \frac{\alpha}{(D-4)},
\end{equation}
and in this singular limit, the effective Einstein-Gauss-Bonnet theory of gravity leads to non-trivial contributions in the gravitational dynamics preserving the number of degrees of freedom and remains free from the Ostrogradsky instability \cite{Glavan:2019inb}.  Further,  this extension of  Einstein's gravity bypasses  all conditions of Lovelock's theorem \cite{Lovelock:1972vz}, and is also free from the singularity problem. 
This $4D$ novel Gauss-Bonnet theory of gravity already attracted several researchers and the charged counterpart of spherically symmetric black holes were also found \cite{Fernandes:2020rpa}.  Among other probes in the theory include discussion on the  the innermost stable circular orbit (ISCO) \cite{Guo:2020zmf},  black hole stability and quasi-normal modes \cite{Konoplya:2020bxa}, construction of rotating counterparts to discuss black holes shadows \cite{Wei:2020ght,Kumar:2020owy}, and also  extension of the theory to more general settings \cite{Konoplya:2020qqh}. 

It is the purpose of this letter to obtain  a class of nonstatic solutions describing radiating
black holes in the  $4D$ novel Gauss-Bonnet theory of gravity. We  discuss how higher  order curvature
corrections  alter black hole solutions and their qualitative features that we know from our knowledge of  black holes in GR. Vaidya \cite{pc} discovered the radiating black hole which is a solution of Einstein's equations with spherically symmetric radially propagating null fluid.  The  Vaidya \cite{pc} geometry offers a more realistic background than static geometries, where all back reactions are ignored (see, Refs.~\cite{ww} for Vaidya solutions in GR  and 
\cite{sgad} for  higher-dimensional Vaidya-like solutions).   Also, several
 Vaidya-like spherical radiating black hole solutions have been explored in Einstein-Gauss-Bonnet gravity
\cite{egb-vaidya,gallo}.   The Vaidya metric has been widely utilized for various 
purposes:  (i) as a testing ground for the
Cosmic Censorship Conjecture (CCC) \cite{Penrose:1964wq}, (ii) as an exterior solution in gravitational collapse  models with heat conducting matter \cite{Herrera:1997ec}, (iii) to discuss Hawking radiation and the  black hole evaporation \cite{Parentani:2000ts}, and also in the (iv)
stochastic gravity program \cite{hv}. 

 The family of solutions discussed here belongs to a type
II fluid. However, when the matter field degenerates to a type I
fluid, we can regain  the static black hole solutions \cite{Glavan:2019inb}. In
particular, our results in the limit $\alpha \rightarrow 0$  generate models \emph{vis-$\grave{a}$-vis}
$4D$ relativistic solutions \cite{adsg}.

\section{Black holes in the novel 4D Einstein-Gauss-Bonnet gravity}
We begin discussing a  static spherically symmetric black hole in the $4D$ novel  Einstein-Gauss-Bonnet gravity \cite{Glavan:2019inb}, where the action for the theory is Eq.~(\ref{action}) with the coupling constant $\alpha$ rescaled to $\alpha/(D-4)$.  For  maximally symmetric spacetimes, the Riemann tensor reads
\begin{eqnarray}
\car^{ab}_{\; \;\; cd}={\frac{\cal K}{D-1}}\left(\delta^a_c  \delta^b_d - \delta^a_d \delta^b_c\right),
\end{eqnarray}
the variation of the Gauss-Bonnet term in this case evaluates to
\begin{equation}\label{gbc}
\frac{g_{bc}}{\sqrt{-g}} \frac{\delta {\cal L}_{GB}}{\delta g_{ac}} = \frac{\alpha (D-2) (D-3)}{2(D-1)} {\cal K}^2 \delta_b^a.
\end{equation}
Obviously, because  of the rescaled Gauss-Bonnet coupling $\alpha/(D-4)$, the variation of the Gauss-Bonnet action does not vanish in $D=4$.  In the limit $D \rightarrow 4$, we obtain two branches of  solution as \cite{Glavan:2019inb}
\begin{eqnarray}
&&{\cal K}_{\pm}=-\frac{3}{4 \alpha}\left[1\pm
\sqrt{1+\frac{8\alpha\Lambda}{3}} \right].
\end{eqnarray}
Thus the novel  Einstein-Gauss-Bonnet gravity admits both branches of the solution in $4D$. Further, it admits spherically symmetric black holes generalizing the  Schwarzschild black holes and can have two horizons depending on the critical mass \cite{Glavan:2019inb}.  The solution, in the $r\to \infty$ limit, the ``$-$" branch leads to the GR Schwarzschild black hole, whereas  the ``$+$" branch gives the Schwarzschild-de Sitter model \cite{Glavan:2019inb,Fernandes:2020rpa}.  Henceforth, we shall restrict ourselves to the ``$-$''  solution.   The solution in  \cite{Glavan:2019inb} actually was also found earlier in the gravity with a conformal anomaly \cite{Cai:2009ua}.

\subsection{Radiating black holes}
The main aim of this work is to find a nonstatic spherically symmetric spacetime, i.e.,  derive the Vaidya-like radiating black holes  the $4D$ novel Einstein-Gauss-Bonnet gravity for the null dust as source  having energy momentum tensor 
\begin{equation}\label{emtn}
T_{a b}= \psi(v,r) \beta_a \beta_b,
\end{equation}
with $\psi(v,r)$ being the nonzero energy density and $\beta_a$ is a null
vector such that $\beta_{a} = \delta_a^0, \beta_{a}\beta^{a} = 0.$ Expressed in terms of Eddington coordinates, the metric of general
spherically symmetric spacetime in
$D$-dimensions \cite{adsg} is given by
\begin{equation}
ds^2 = - A(v,r)^2 f(v,r)\;  dv^2
+  2 \epsilon A(v,r)\; dv\; dr + r^2\gamma_{ij} dx^idx^j,
\label{eq:me2}
\end{equation}
where $\{ x^a \} = \{ v,\;r,\; \theta_1, \ldots \; \theta_{D-2} \}$.
For null dust, $T_{vr}$ must be non-zero and $T^v_v=T^r_r$ for
the null energy condition.  Using $T^v_r=0$, we get $A(v,r)=A(v)$ which could be
set as $1$  by redefining time.  It is useful to introduce
a local mass function $m(v,r)$ defined by \[ f(v,r) = 1 - \frac{2
	m(v,r)}{(D-3)r^{(D-3)}}. \] For $m(v,r) = M(v)$ and $A=1$,
the metric reduces to the $D$-dimensional Vaidya metric \cite{ns,adsg}.   Let us consider the metric (\ref{eq:me2}) with stress tensor (\ref{emtn}) and apply the procedure in \cite{Glavan:2019inb,Fernandes:2020rpa}.  Now, in the limit $D\to 4$, the $(r,r)$ equation of (\ref{field_eq}) reduces to
\begin{equation}\label{rreq}
\Big[r^2+2\alpha (1-f)\Big]\frac{1}{r^3} \frac{\partial f}{\partial r}  -2\Big[r^2-\alpha (1-f)\Big]\frac{1-f}{2r^4}=0. 
\end{equation}
Solving Eq.~(\ref{rreq}), we obtain
\begin{eqnarray}\label{vegb}
f(v,r)=1+
\frac{r^2}{4 \alpha}
\left[1 \pm \sqrt{1
	+\frac{ 16\alpha M(v)}{r^{3}}}
\right],
\end{eqnarray}
where $M(v)$ is positive and an arbitrary function of $v$ identified
as mass of the matter. The special case in which $\dot{M}(v) = 0 $, Eq.~(\ref{vegb}) leads to the Schwarzschild
solution \cite{Glavan:2019inb} of the theory in  Eddington-Finkelstein coordinates. 
This metric is indeed a solution of the field equations~(\ref{field_eq})
and  again, there are two families of solutions which
correspond to the sign in front of the square root in
Eq.~(\ref{vegb}).   In the GR limit ${ \alpha} \to 0$, the minus-branch solution reduces to the Vaidya solution \cite{pc,sgad}. 
 From $T^r_v = G^r_v$, in the limit $D \to 4$, we obtain the energy density of the null fluid as
\begin{equation}\label{energy}
\psi(v,r) = \frac{1}{r^2} \frac{dM}{dv}.
\end{equation}
The family of solutions discussed here, in general, belongs to a Type II fluid defined in \cite{he}. When
$M(v)=M$=const., we have $\psi$=0, and the matter field degenerates to a Type I fluid \cite{ww}, and   we can
generate static black hole solutions obtained in Ref.~ \cite{Glavan:2019inb}. Indeed
the static black hole solutions in  Eddington-Finkelstein
coordinates can
be recovered by setting $M(v) = M$, with $ M $ as constant, in which case $f(v,r) \rightarrow f(r)$. In the
static limit, we can obtain from the metric (\ref{eq:me2}), the usual form by means of the coordinate transformation
\begin{equation}
dv = A(r)^{-1} \left( dt + \epsilon \frac{dr}{f(r)} \right).
\end{equation}
In the case of spherical symmetry, even when $f(r)$ is replaced by
$f(t,r)$, we can cast the metric in the form (\ref{eq:me2})
\cite{visser}.

\paragraph{Energy conditions:}
 The weak energy condition (WEC) demands that the energy
momentum tensor obeys $T_{ab}w^a w^b \geq 0$ for any
timelike vector, i.e., while the strong energy condition (SEC)
holds for Type II fluid if WEC is true, i.e., both WEC and SEC,
for a Type II fluid, are identical \cite{ww,he}. \\
 The dominant energy conditions (DEC) holds when for a 
timelike vector $w_a$, $T^{ab}w_a w_b \geq 0$, and $T^{ab}w_a$ is a
non-spacelike vector.  This, in general, is satisfied if  $\psi (v,r)
> 0$ giving a restriction on the choice of the function $M(v)$.   

 Since $T^r_v $ is the only non-zero component, from Eq.~(\ref{energy}),  we observe $\psi(v,r) > 0$ requires $\dot{M(v)} >0$  and all the energy conditions are obeyed. 

\section{Structure of the horizons}
In this section, we discuss the structure and location of three black hole surfaces viz., time-like limit surface ($ TLS $), event horizon  ($ EH $) and apparent horizon ($ AH $)  of the radiating black hole (\ref{vegb}) in the $4D$ novel Einstein-Gauss-Bonnet gravity, and compare with the GR case.  It turns out that for the static spherically symmetric Schwarzschild black hole (which does not radiate), the three horizons degenerate to $r=2M$.  While for nonstatic Vaidya black hole, with small luminosity $L_M$,  where  spherical symmetry is still respected, one has $TLS=AH$, but $EH \neq AH$.  The mass of the black hole is defined by $M(v)$ and the luminosity due to loss of mass is given by $L_M \approx - dM/dv$, $L_M < 1$ measured in the region where $d/dv$ is timelike \cite{jygb}. 

Let us suppose that $v=$ const. is a ingoing surface with null tangent vector (normal) $l^a$, the metric of $v=$ const. will degenerate to the $2D$ surface $\gamma_{ab}$ \cite{jygb}
\begin{equation}\label{gabgb}
\gamma_{ab} = r^2 \delta_a^{\theta} \delta_b^{\theta} + r^2
\sin^2 \theta \delta_a^{\varphi} \delta_b^{\varphi},
\end{equation}
and also we define the outgoing null geodesics by the tangent vector $\beta^{a}$ such that 
\begin{eqnarray}
\beta_{a} &=& - \delta_a^v, \: l_{a} = - \frac{1}{2} f(v,r)
\delta_{a}^v + \delta_a^r, \label{nvagb} \\
l_{a}l^{a} &=& \beta_{a} \beta^{a} = 0, \; ~l_a \beta^a = -1,\;
\;l^{a}\gamma_{ab} = 0, \nonumber\\&& ~\gamma_{ab}\; ~\beta^{b} = 0,
\label{nvdgb}
\end{eqnarray}
with $f(v, r)$ given by Eq.~(\ref{vegb}).  Then one has that 
 \begin{equation}\label{deco}
g_{ab} = \gamma_{ab} - l_{a} \beta_{b} - \beta_a l_{b}.
 \end{equation}
The optical behavior
of null geodesics congruences is governed by the Raychaudhuri
equation \cite{Ghosh:2008jca,mrm,rm}
\begin{equation}\label{regb}
\frac{d \Theta}{d v} = K \Theta - R_{ab}l^al^b-\frac{1}{2}
\Theta^2 - \sigma_{ab} \sigma^{ab} + \omega_{ab}\omega^{ab},
\end{equation}
with expansion $\Theta$, twist $\omega$, shear $\sigma$, and surface
gravity $K$. Here we are interested in the outgoing null geodesics. The expansion of the null rays parameterized by $v$ is
given by
\begin{equation}\label{theta}
\Theta = \nabla_a l^a - K,
\end{equation}
where the $\nabla$ is the covariant derivative and on the horizon the surface
gravity can be calculated using 
\begin{equation}\label{sggb}
K = - \beta^a l^b \nabla_b l_a.
\end{equation}
If $v$ is the time parameter and $\lambda$ is affine parameter related to $v$  by $K = \ddot{\lambda} (\dot{\lambda}^{-1})$. 
The $ TLS $ can be obtained by solving $g_{vv}=0$, which gives 
\begin{equation}\label{tls}
r^{TLS} = M(v) + \sqrt{M(v)^2 - 2 \alpha }.
\end{equation}
 The $ AH $ can be either null or
spacelike and are defined as surfaces such that $\Theta \simeq 0$
which implies that $f=0$ \cite{jygb}. Using Eqs.~(\ref{nvagb}) and (\ref{sggb})
\begin{equation}
K = \frac{r}{4 \alpha} \left[1 - \sqrt{1 + 16 \alpha \frac{M(v)}{r^3}
}\right] - \frac{\frac{3M(v)}{r^2} }{\sqrt{1 + 16 \alpha
		\frac{M(v)}{r^4} }}.\label{Kgb}
\end{equation}
Then Eqs.~(\ref{nvagb}),
(\ref{theta}),  and  (\ref{Kgb}) yield the expansion parameter
\begin{equation}
\Theta = \frac{1}{r} \left[1 + \frac{r^2}{4\alpha}\left[1 - \sqrt{1
	+ 16 \alpha \left(\frac{M(v)}{r^3}\right)}\right]\right]. \label{thgb}
\end{equation}
From the Eq.~(\ref{thgb}) it is clear that $AH$ is 
\begin{equation}\label{aegb}
r_{AH} = M(v) \pm \sqrt{M(v) ^2- 2 \alpha},
\end{equation}
which means  that the $ AH $ is the outermost marginally trapped
surface for the outgoing photons due to the fact that $\Theta \approx 0$ at $r=r_{AH}$. Further, we observe that $r_{TLS}  =  r_{AH}$ and the $AH$ is also a timelike surface, for
 $\alpha \rightarrow 0$ then $r_{AH}
=2 M(v)$. Hence, our solution reduces to the solution in Refs.~\cite{mrm, Ghosh:2008jca} in $4D$  space-time. The Hawking temperature $T_{AH}$ can be determined through the relation 
$ T_{AH} = {K}/{2\pi} $. 

 For an outgoing null fluid, geodesics must obey the null condition
\begin{equation}
\dot{r} = \frac{dr}{dv} = \frac{1}{2}\left[1 + \frac{r^2}{4\alpha}\left[1 -
\sqrt{1 + 16 \alpha \left(\frac{M(v)}{r^3}\right)}\right]\right].
\label{rnggb}
\end{equation}   
which means an outgoing radial null geodesic satisfy (\ref{rnggb}). Differentiating $r$ with respect to $v$, we obtain
\begin{equation}\label{rdd}
\ddot{r} = \frac{r\dot{r}}{4\alpha}\left(1 - \sqrt{1 + 16\alpha
	\frac{M(v)}{r^3}}\right) + \frac{\frac{L_M}{r} +  \frac{3
		M(v)\dot{r}}{r^2}}{\sqrt{1 + 16\alpha\frac{M(v)}{r^4}}}.
\end{equation}
At the timelike surface $r_{AH}$, one has $\dot{r}=0$ and hence 
 $\ddot{r}> 0$ for $L >0$.  Hence the photon will escape from the $r_{AH}$ and can reach arbitrarily large distance.  This means that $r_{AH}$ cannot be $EH$.  It means that the photon will stay briefly at this surface. 

The $EH$ is a null three-surface defined by the locus of outgoing
future-directed null geodesic rays that never manage to reach
arbitrarily large $r$.  It means that  the photons at $EH$ are not accelerated thereby
\begin{equation}
\left[\frac{d^2r}{dv^2}\right]_{{EH}} \simeq ~ 0. \label{ehgb}
\end{equation}
Then Eqs.~(\ref{Kgb}) and (\ref{thgb}) can be used to put
Eq.~(\ref{ehgb}) in the form
\begin{equation}
K \Theta_{EH} \simeq  \left[ \frac{1}{r^2}\frac{\partial f}{\partial
	v} \right]_{EH} \simeq  - \frac{f(v,r)}{{r_{EH}^2}} \frac{{L_M}}{\sqrt{1
		+ 16 \alpha \left(\frac{M(v)}{r_{EH}^4}\right)}},  \label{eh1gb}
\end{equation}
where the expansion is
\begin{equation}
\Theta_{EH}\simeq \frac{1}{ r_{EH}}\left[1 +
\frac{r^2_{EH}}{4\alpha}\left[1 - \sqrt{1 + 16\alpha
	\left(\frac{M(v)}{r^3_{EH}}\right)}\right]\right]. \label{thgb1}
\end{equation}
For the null vectors $l_a$ in Eq.~(\ref{nvagb}) and the component of
energy momentum tensor yields
\begin{equation}
R_{a b}l^{a}l^{b} =  \frac{3}{2r} \frac{\partial f}{\partial v}.
\label{chigb}
\end{equation}
The Raychaudhuri equation,  for the spherical symmetric case ($\sigma = \omega = 0$), yields  \cite{jygb}:
\begin{equation}
\frac{d \Theta}{d v} = K \Theta - R_{ab}l^al^b- \frac{1}{2}
\Theta^2.  \label{mregb}
\end{equation}
Thus neglecting $\Theta^2$, Eqs.~(\ref{eh1gb}), (\ref{chigb}) and
(\ref{mregb}), imply that
\begin{equation}
\left[ \frac{d \Theta}{d v} \right]_{EH} \simeq 0.\label{eh}
\end{equation}
The $EH$ in our case are therefore placed by
Eq.~(\ref{eh}), which admits  the solution 
\begin{equation}
r_{EH} =M^{*}(v) \pm \sqrt{M^{*}(v) ^2- 2 \alpha},
\end{equation}
where
\begin{equation}
M^{*}(v) = M(v) - \frac{L_M}{K}.
\end{equation}
 The region between the $AH$ and the $EH$, $r_{EH}<r< r_{AH}$, is defined as a \emph{quantum ergosphere},  which  does not exist for a static black hole \cite{jygb}. In the ergosphere, 
photons are locally trapped but, being outside the $EH$, they can cross the $AH$ at a later
stage  and propagate to infinity. 
Thus, because of Hawking evaporation, for
the static Schwarzschild black hole $r_{EH}=r_{AH}$.  The results presented, in the limit $\alpha \rightarrow 0$, go over to that of $4D$ Vaidya solutions \cite{jygb}. 

\section{CONCLUSION}
 Einstein-Gauss-Bonnet gravity is a natural extension of  GR to higher dimensions in which the first and  second  terms in the  action correspond, respectively, to the  Ricci scalar and curvature squared Gauss-Bonnet term. It has several  additional nice properties than Einstein's GR \cite{Deser:2002jk}, but it is topological in $4D$ and does not make a contribution to the gravitational dynamics.   However in the $4D$ novel  Einstein-Gauss-Bonnet gravity \cite{Glavan:2019inb}  the Gauss-Bonnet coupling is scaled to $\alpha/(D-4)$, leading to a non-trivial contribution in $4D$ spacetimes if we take the limit $ D \rightarrow 4 $ while finding equations of motion.  

We have found exact Vaidya-like radiating black hole solutions in the $4D$  novel Einstein-Gauss-Bonnet gravity which are characterized by the mass $ M(v) $ and  the parameter $\alpha$.  We have shown that a radiating black hole  has three  horizon-like loci, viz. $ AH $, $ EH  $ and $ TLS $ with  $r_{EH} < r_{AH} = r_{TLS}$. We showed that the effect of the coupling constant  $\alpha$  on the structure and location of these three horizon surfaces,  which are changed when compared with analogous GR case, is significant in the dynamical evolution of the black hole horizons with the effect of higher order curvature in $4D$.  

The lack of exact solutions that are suitable to study gravitational collapse makes progress very difficult  in studying CCC as we are still far away from its proof.  The  Penrose \cite{Penrose:1964wq,rp} CCC, in its weak version,  essentially states  that any naked singularity which is created by evolution of regular initial data will be shielded from the external view by an $ EH $. According to the strong version of the CCC, naked singularities are never produced.  Our  solutions of the  $4D$  novel Einstein-Gauss-Bonnet gravity are dynamical which can be useful to get insights into a more general gravitational collapse setting.  Hence  it would be interesting to consider gravitations collapse in this higher curvature  gravity in realistic $4D$
 spacetimes which is being considered.  Such studies will help to formulate  the CCC in a precise mathematical form.

Further, the results presented here are a generalization of previous discussions, of radiating black holes of GR, to a more general setting. The possibility of  generalization of these results to  include rotation and to more general Lovelock gravity theories  \cite{Konoplya:2020qqh}  are interesting problems for future research.

\section*{Acknowledgments}
Authors would like to thank DST INDO-SA bilateral project DST/INT/South Africa/P-06/2016, S.G.G. also thank SERB-DST for the ASEAN project IMRC/AISTDF/CRD/2018/000042 and Rahul Kumar for fruitful discussions. S.D.M. acknowledges
that this work is based upon research supported by the South African
Research Chair Initiative of the Department of Science and
Technology and the National Research Foundation.

\appendix
\appendix
\section{Charged radiating AdS black holes}
Several extensions of Vaidya solutions in which the source is a mixture of a
perfect fluid and null radiation have been obtained in later years
\cite{ka}. This includes the Bonnor-Vaidya solution \cite{bv} for
the charged  case; when we extend to the novel 4D Einstein-Gauss-Bonnet gravity, the solution is given by
\begin{eqnarray}\label{cvegb}
f_c(r, v)=1+
\frac{r^2}{4 \alpha}
\left[1 - \sqrt{1
	+16\alpha \left( \frac{M(v)}{r^{3}} - \frac{q(v)^2}{2 r^4} - \frac{1}{2 l^2}\right) }
\right].
\end{eqnarray}
 The minus-branch of solution (\ref{cvegb}), in the GR limit ${ \alpha} \to 0$, recovers the Bonnor-Vaidya solution \cite{bv}.  
The luminosity due to loss of mass is given by
$L_M = - dM/dv$, $L_M < 1$ , and  similarly due to  charge by $ L_q = -
dq/dv$, where $L_M,  L_q < 1$, both measured in the  region
where $d/dv$ is timelike \cite{jygb}.  In order to further discuss the physical
nature of our solutions, as above, we calculate the their kinematical parameters with vanishing cosmological constant.  
We obtain the  surface gravity
\begin{eqnarray}\label{kgbc}
K && = \frac{r}{4 \alpha}  	\left[1 - \sqrt{1
	+16\alpha \left( \frac{M(v)}{r^{3}} - \frac{q(v)^2}{2 r^4} - \frac{1}{2 l^2}\right) }
\right] \nonumber \\
& & +  \frac{ \left(\frac{3M(v)}{r^2} - \frac{2 q(v)^2}{r^3}\right)}{\sqrt{1
		+16\alpha \left( \frac{M(v)}{r^{3}} - \frac{2 q(v)^2}{r^4} - \frac{1}{2l^2}\right). }} 
\end{eqnarray}
Then the expansion of null ray congruence becomes
\begin{equation}\label{thgbc}
\Theta =\frac{1}{r}\left[ 1+
\frac{r^2}{4 \alpha}
\left[1 - \sqrt{1
	+16\alpha \left( \frac{M(v)}{r^{3}} - \frac{q(v)^2}{2 r^4} - \frac{1}{2l ^2}\right) }
\right]\right]. 
\end{equation}
The $ AHs$ are defined as surfaces such that $\Theta \simeq 0$  and for $1/l^2=0$ and are given by
\begin{equation}\label{CAH}
r_{AH} = M(v) \pm \sqrt{M(v)^2 - q(v)^2 - 2 \alpha}
\end{equation}
  Here one sees that $\Theta
= 0$ implies $f=0$, and also  $g_{vv}(r=r_{AH}) = 0$ implies that
$ AH = TLS $ in our non-rotational case.
As above, the $ EH $ are strictly
null, and are defined to order of $O(L_M,  L_q)$ \cite{jygb}.   The requirement  for acceleration of
null-geodesic congruences at the $ EH $ is given in Eq.~(\ref{ehgb}).  
An outgoing radial null geodesic which is
parameterized by $v$  satisfies
\begin{equation}\label{ongc}
\frac{dr}{dv} = 1+
\frac{r^2}{4 \alpha}
\left[1 - \sqrt{1
	+16\alpha \left( \frac{M(v)}{r^{3}} - \frac{q(v)^2}{2 r^4} - \frac{1}{2 l^2}\right) }
\right],
\end{equation}
and 
\begin{equation}\label{kthc}
K \Theta_{EH} = \frac{\frac{L_M}{r_{EH}^2}- \frac{q(v) L_q}{r_{EH}^3}}{ \sqrt{1
		+16\alpha \left( \frac{M(v)}{r_{EH}^3} - \frac{q(v)^2}{r_{EH}^4} - \frac{1}{l^2}\right) }},
\end{equation}
where the expansion is
$ \Theta_{EH}\simeq  f_c(v,r_EH)/{(2 r_{EH}})$. 
The using Eqs. (\ref{chigb}) and (\ref{mregb})  we find that EH is given by (\ref{CAH}) 
with $M$ and $q$ being respectively replaced by $M^*$ and $q^*$
\cite{sgplb}, where $M^*$ and $q^*$ are the effective mass and charge
defined as follows
\begin{equation}
M^{*}(v) = M(v) - \frac{L_M}{\kappa},q^{*}(v) = q(v) -
\frac{ L_q}{\kappa}.
\end{equation}
The results in this section, for $q(v)=0$, go over to the results presented in the main paper.  In the limit $\alpha \to 0$ they coincide with that of the Bonnor-Vaidya solutions.


\begin{thebibliography}{99}
\bibitem {dll}D. Lovelock, J. Math. Phys. (N.Y.) \textbf{12}, 498 (1971).


\bibitem{Lanczos:1938sf}
C.~Lanczos,
Annals Math.\  {\bf 39} 842 (1938).



\bibitem{Gross}   D. J.~Gross and E.~Witten,   Nucl. Phys. {B 277}, 1 (1986);
D. J.~Gross and J. H.~Sloan,   Nucl. Phys. B  {291}, 41 (1987).

\bibitem{Boulware:1985wk}  
D. G.~Boulware and S.~Deser, Phys. Rev. Lett. { 55}, 2656 (1985); 
J. T.~Wheeler, Nucl. Phys. B {\bf 268}, 737 (1986).


\bibitem{Wiltshire:1985us} 
D.~L.~Wiltshire,
Phys.\ Lett.\ B {\bf 169}, 36 (1986).

\bibitem {egb}Y. M. Cho and I. P. Neupane, Phys. Rev. D { 66}, 024044
(2002); S.~Nojiri and S.~D.~Odintsov,
Phys.\ Lett.\ B {\bf 521}, 87 (2001);
Erratum: [Phys.\ Lett.\ B {\bf 542}, 301 (2002)]; S.~Nojiri, S.~D.~Odintsov and S.~Ogushi,
Phys.\ Rev.\ D {\bf 65}, 023521 (2002);
R. G. Cai, Phys. Rev. D {\bf 65}, 084014 (2002); 
M. Cvetic, S. Nojiri and S. D. Odintsov, Nucl. Phys. B {628}, 295 (2002);    
I. P. Neupane, Phys. Rev. D {\bf 67}, 061501(R) (2003); {\bf 69}, 084011 (2004);  
N. Deruelle, J. Katz, and S. Ogushi, Class Quant. Grav.  {\bf 21}, 1971 (2004); 
R. G. Cai and Q. Guo, Phys. Rev. D {\bf 69}, 104025 (2004); 
T. Torii and H. Maeda, Phys. Rev. D {\bf 71}, 124002 (2005); 
M. H. Dehghani and R. B. Mann, Phys. Rev. D {\bf 72}, 124006 (2005); 
M. H. Dehghani and S. H. Hendi, Phys. Rev. D {\bf 73}, 084021 (2006);
E.~Herscovich and M.~G.~Richarte,
Phys.\ Lett.\ B {\bf 689}, 192 (2010).


\bibitem{egb2} 
S.~Mignemi and N.~R.~Stewart,
Phys.\ Rev.\ D {\bf 47}, 5259 (1993);
S.~O.~Alexeev and M.~V.~Pomazanov,
Phys.\ Rev.\ D {\bf 55}, 2110 (1997);
T.~Torii, H.~Yajima and K.~I.~Maeda,
Phys.\ Rev.\ D {\bf 55}, 739 (1997);
R.~Konoplya,
Phys.\ Rev.\ D {\bf 71}, 024038 (2005);
S.~Jhingan and S.~G.~Ghosh,
Phys.\ Rev.\ D {\bf 81}, 024010 (2010);
B.~Kleihaus, J.~Kunz and E.~Radu,
Phys.\ Rev.\ Lett.\  {\bf 106}, 151104 (2011);
A.~Maselli, P.~Pani, L.~Gualtieri and V.~Ferrari,
Phys.\ Rev.\ D {\bf 92}, 083014 (2015).
\bibitem{Kanti:1996gs}
P.~Kanti and K.~Tamvakis,
Phys. Lett. B \textbf{392}, 30 (1997).

\bibitem{Antoniou:2017hxj}
G.~Antoniou, A.~Bakopoulos and P.~Kanti,
Phys. Rev. D \textbf{97}, 084037 (2018).

\bibitem{Antoniou:2017acq}
G.~Antoniou, A.~Bakopoulos and P.~Kanti,
Phys. Rev. Lett. \textbf{120}, 131102 (2018).

\bibitem{Bakopoulos:2018nui}
A.~Bakopoulos, G.~Antoniou and P.~Kanti,
Phys. Rev. D \textbf{99},  064003 (2019).

\bibitem{Ghosh1:2018bxg} 
S.~G.~Ghosh, D.~V.~Singh and S.~D.~Maharaj,
Phys.\ Rev.\ D {\bf 97}, 104050 (2018);
S.~Hyun and C.~H.~Nam,
Eur.\ Phys.\ J.\ C {\bf 79}, 737 (2019);
A.~Kumar, D.~Veer Singh and S.~G.~Ghosh,
Eur.\ Phys.\ J.\ C {\bf 79}, 275 (2019);
D.~V.~Singh, S.~G.~Ghosh and S.~D.~Maharaj,
Annals Phys.\  {\bf 412}, 168025 (2020).

\bibitem{Glavan:2019inb}
D.~Glavan and C.~Lin,
Phys.\ Rev.\ Lett.\  {\bf 124}, 081301 (2020).

\bibitem{Cai:2009ua} 
R.~G.~Cai, L.~M.~Cao and N.~Ohta,
JHEP {\bf 1004}, 082 (2010); R.~G.~Cai,
Phys.\ Lett.\ B {\bf 733}, 183 (2014). 

\bibitem{Lovelock:1972vz}
D.~Lovelock,
J.\ Math.\ Phys.\  {\bf 13} 874 (1972).


\bibitem{Fernandes:2020rpa} 
P.~G.~S.~Fernandes,
arXiv:2003.05491 [gr-qc].


\bibitem{Guo:2020zmf} 
M.~Guo and P.~C.~Li,
arXiv:2003.02523 [gr-qc].

\bibitem{Konoplya:2020bxa}
R.~A.~Konoplya and A.~F.~Zinhailo,
arXiv:2003.01188 [gr-qc].


\bibitem{Wei:2020ght} 
S.~W.~Wei and Y.~X.~Liu,
arXiv:2003.07769 [gr-qc].

\bibitem{Kumar:2020owy} 
R.~Kumar and S.~G.~Ghosh,
arXiv:2003.08927 [gr-qc].


\bibitem{Konoplya:2020qqh} 
R.~A.~Konoplya and A.~Zhidenko,
arXiv:2003.07788 [gr-qc].

\bibitem{pc} 
P. C.~Vaidya, { Proc. Indian Acad. Sci.} {\bf A33}, 264 (1951);
P. C.~Vaidya, {Gen. Relativ. Grav.} {\bf 31}, 119 (1999).

\bibitem{ww} A.~Wang and Y.~Wu, {Gen. Relativ. Grav.} {\bf31}, 107 (1999).

\bibitem{sgad}  S.~G.~Ghosh and D.~Kothawala, Gen.\ Relativ.\ Gravit.\  {\bf 40}, 9 (2008).

\bibitem{gallo} A. E. Dominguez and E. Gallo, Phys. Rev. D \textbf{73}, 064018 (2006).

\bibitem{egb-vaidya} 
T. Kobayashi, {Gen. Relativ. Grav.} {\bf 37}, 1869 (2005);
H. Maeda, Class. Quantum Grav. \textbf{23}, 2155 (2006);
S.~G.~Ghosh and N.~Dadhich,
Phys.\ Rev.\ D {\bf 82}, 044038 (2010);
S.~G.~Ghosh and S.~D.~Maharaj,
Phys.\ Rev.\ D {\bf 89}, 084027 (2014);
S.~G.~Ghosh and S.~D.~Maharaj,
Eur.\ Phys.\ J.\ C {\bf 75}, 7 (2015).

\bibitem{sgplb} S.~G.~Ghosh,
Phys.\ Lett.\ B {\bf 704}, 5 (2011); 


\bibitem{Penrose:1964wq} 
R.~Penrose,
Phys.\ Rev.\ Lett.\  {\bf 14}, 57 (1965).


\bibitem{Herrera:1997ec} 
L.~Herrera and J.~Martinez,
Gen.\ Relativ.\ Grav.\  {\bf 30}, 445 (1998).


\bibitem{Parentani:2000ts} 
R.~Parentani,
Phys.\ Rev.\ D {\bf 63}, 041503 (2001).


\bibitem{hv} B. L. Hu and  E.~Verdaguer, Living Rev. Relativity {\bf 7}, 3 (2004).


\bibitem{adsg}  D.~Kothawala and S.~G.~Ghosh, Phys.\ Rev.\  D {\bf 70}, 104010 (2004).

\bibitem{ns}S.~G.~Ghosh and N.~Dadhich,
Phys.\ Rev.\ D {\bf 64}, 047501 (2001).

\bibitem{he} S. W.~Hawking and G. F. R.~Ellis, {\it The Large Scale Structure of
	Space-time} (Cambridge University Press, Cambridge,  1973).

\bibitem{visser} A. B.~Nielsen and M.~Visser, Class. Quantum Grav. {\bf 23}, 4637 (2006).

\bibitem{jygb} J. W.~York, Jr., in {\it Quantum Theory of Gravity:
	Essays in Honor of Sixtieth Birthday of Bryce S. DeWitt}, edited by
S.Christensen (Hilger, Bristol, 1984), p.135.

\bibitem{Ghosh:2008jca} S.~G.~Ghosh and D.~W.~Deshkar,
Phys.\ Rev.\ D {\bf 77}, 047504 (2008).

\bibitem{rm} R. L. Mallet, {Phys. Rev. D} {\bf 33}, 2201 (1986);
B. D.~Koberlein and R. L.~Mallet, {Phys. Rev. D} {\bf 49}, 5111 (1994).

\bibitem{mrm} M. R. Mbonye, {Phys. Rev. D} {\bf 60}, 124007 (1999).

\bibitem{Deser:2002jk}
S.~Deser and B.~Tekin,
Phys.\ Rev.\ D {\bf 67}, 084009 (2003); J.T. Wheeler, Nucl. Phys. \textbf{B 268}, 737 (1986).


\bibitem{rp} R.~Penrose, {\it Riv Nuovo Cimento} {\bf 1}, 252 (1969); {\it General Relativity}, {\it an Einstein Centenary
	Volume}, edited by S. W.~Hawking and W.~Israel (Cambridge
University Press, Cambridge) .

\bibitem{ka} ~Krasiaski, A. (1997). {\it Inhomogeneous Cosmological Models}
(Cambridge University Press, Cambridge).

\bibitem{bv} W. B.~Bonnor  and P.C.~Vaidya,  Gen. Relativ, Gravit.
{\bf 1}, 159 (1970).






\end{thebibliography}
\end{document}